\documentclass[fleqn,10pt]{wlscirep}
\usepackage[utf8]{inputenc}
\usepackage[T1]{fontenc}
\usepackage{float}
\usepackage{soul}

\usepackage[justification=justified]{caption}
\graphicspath{ {Graphics/} }
\title{Experimental Characterization of Ultrafast, Tunable and Broadband Optical Kerr Nonlinearity in Graphene}

\author[1,2]{Siddharatha Thakur}
\author[1,2,3]{Behrooz Semnani}
\author[1]{Safieddin Safavi-Naeini}
\author[1,2,4*]{Amir Hamed Majedi}

\affil[1]{Department of Electrical \& Computer Engineering, University of Waterloo, Waterloo, N2L3G1, Canada}
\affil[2]{Waterloo Institute for Nanotechnology, University of Waterloo, Waterloo, N2L3G1, Canada}
\affil[3]{Institute for Quantum Computing, University of Waterloo, Waterloo, N2L3G1, Canada}
\affil[4]{Department of Physics and Astronomy, University of Waterloo, Waterloo, N2L3G1, Canada}

\affil[*]{ahmajedi@uwaterloo.ca}

\begin{abstract}
In this study we systematically measure the near-infrared spectral dependence, the sub-picosecond temporal evolution and pulse-width dependence of the effective Kerr coefficient ($n_{2,eff}$) of graphene in hundreds of femtosecond regime. The spectral dependence measured using the Z-scan technique is corroborated by quantum theory to extract a $n_{2,eff} \propto \lambda^2$ dependence. The temporal evolution extracted using the time-resolved Z-scan measurement shows the nonlinear response peaking at zero delay time and relaxing on a time scale of carrier relaxation. Since the Kerr-type response originated from the optically induced carrier population difference, the time-scale of the evolution of the nonlinear response is apt. The $n_{2,eff}$ shows a dependence on the pulse-width attributed to the relative heating and cooling times of the carriers. This dependence is strong when the pulse-duration is on the same time-scale as the decay constant. Throughout our study the $n_{2,eff}$ remains positive.
\end{abstract}

\begin{document}

\flushbottom
\maketitle
% * <john.hammersley@gmail.com> 2015-02-09T12:07:31.197Z:
%
%  Click the title above to edit the author information and abstract
%
\thispagestyle{empty}

% \noindent Please note: Abbreviations should be introduced at the first mention in the main text – no abbreviations lists. Suggested structure of main text (not enforced) is provided below.

\section*{Introduction}

The Dirac band structure of graphene \cite{wallace, novo_geim} has endowed this atomically thin two-dimensional material with exceptional optical and transport properties. As a scale-invariant host of chiral carriers\cite{semnani}, graphene exhibits a universal optical response in absorption, with a single sheet capable of absorbing $\sim 2.3\%$ of normally incident light. Its reduced dimensionality together with the peculiar Dirac type dynamics of the quasipartciles enable graphene to have a relatively strong nonlinear optical response in the presence of intense electromagnetic illuminations \cite{hendry, zhang_2012}. Intrigued by its potential applications in  photonic and optoelectronics several theoretical and experimental studies of the nonlinear optical properties of graphene have been carried out \cite{hendry, krishna, ciesielski,jiang,vermeulen, vermeulen_spm,zhang_2012, chen, miao, demetriou, dremetsika_2016, ahn, soh, ooi, semnani, cheng_2015, mikhailov_2008, zhang_2011, cheng_2014, zhang_2011,chatz}. Due to the centrosymmetric crystalline structure of graphene, even-order nonlinearities are forbidden and the first nonlinear contribution is a third order effect. In particular, the Kerr type nonlinearity is dominantly attributable to the third order nonlinear interactions in graphene. The refractive index, $n$, considering the first nonlinear term is intensity dependent and  given by the expression $n = n_0 + n_2I$, where $n_0$ is the linear refractive index, $n_2$ is the Kerr coefficient, and $I$ is incident intensity. This Kerr-type nonlinearity in graphene has been reported to be large, confirmed by studies characterising frequency mixing \cite{hendry, krishna, ciesielski}, harmonic generation \cite{jiang}, self-phase modulation\cite{vermeulen, vermeulen_spm}, and self-refraction\cite{zhang_2012, chen, miao, demetriou, dremetsika_2016, ahn}. 

Several theoretical works have attempted to provide a cohesive theory for the third order nonlinearity in graphene, addressing the magnitude\cite{soh, ooi, semnani, cheng_2015} of the nonlinear refractive index ($n_2$), the spectral\cite{semnani, hendry, soh} and temporal\cite{mikhailov_2008, zhang_2011} dependence, and effects of Fermi energy modulation\cite{cheng_2014, ooi, semnani, cheng_2015, chatz}. Although topological anomalies in graphene hinder the adaption of perturbation theory in the treatment of the optical response of graphene \cite{semnani, semnani18}, fortunately, over the optical wavelength range, nonlinear response coefficients can still describe interactions of photons. Perturbative treatment of the nonlinear optical response of graphene yields explicit expressions for the higher orders conductivity tensors. Throughout this work we refer to the third order frequency mixing conductivity as $\sigma^{(3)} (\omega_p,\omega_q,\omega_r)$ where $\omega_p$ , $\omega_q$ and $\omega_r$ are the frequencies of the photons undergoing nonlinear interaction through the graphene monolayer. The conductivity is proportional to the Kerr coefficient through the third order susceptibility. The studies show large tunability in $n_2$ through Fermi level modulation. Other factors such as the spectral and temporal dependence have seen relatively less explicit theoretical discussion.

The experimental studies probing the Kerr-nonlinearity in graphene have shown variation of $n_2$ spanning six order of magnitudes (10$^{-12}$  to 10$^{-7}$ cm$^2$/W) \cite{hendry, zhang_2012, chen, dremetsika_2016}. A point of contention while comparing these results are the varying spectral and temporal properties of the excitation source, sample preparation techniques, and substrate material. The measurements are performed with the excitation wavelength ranging from the visible to the mid-infrared, while the pulse duration ranges from hundred femtoseconds to tens of picoseconds, under different sample preparation methods\cite{berciaud}. It has been recognised that the $n_2$ depends on the wavelength, pulse duration and power in semiconductor systems \cite{ogusu}. Particularly in graphene, the $n_2$ expected to become larger with longer wavelengths \cite{hendry}. This trend has been observed and reported thereafter\cite{miao, ahn}. It has also been shown in graphene oxide, that the nonlinear parameters of the system change when excited with nanosecond and picosecond pulses\cite{liu}. The decrease of $n_2$ with increasing incident intensity due to saturation effects has been observed and reported previously in several studies\cite{zhang_2012, ahn, miao, meng}. Raman studies have shown \textit{p}-type doping occurs when chemical vapour deposition (CVD) graphene is transferred onto quartz \cite{berciaud}. Considering these factors, it is more apt to consider the measured nonlinear parameters in conjunction with the experimental conditions. This makes it imperative to understand the dependence of these factors in a systematic manner to gain a fundamental understanding of the governing processes. The integrability of graphene combined with its large Kerr nonlinearity makes it ideal for use in on-chip all-optical communication and signal-processing with large tunability, which is limited in silicon photonics.

\begin{figure}[]
\includegraphics[scale=0.085, width=\linewidth]{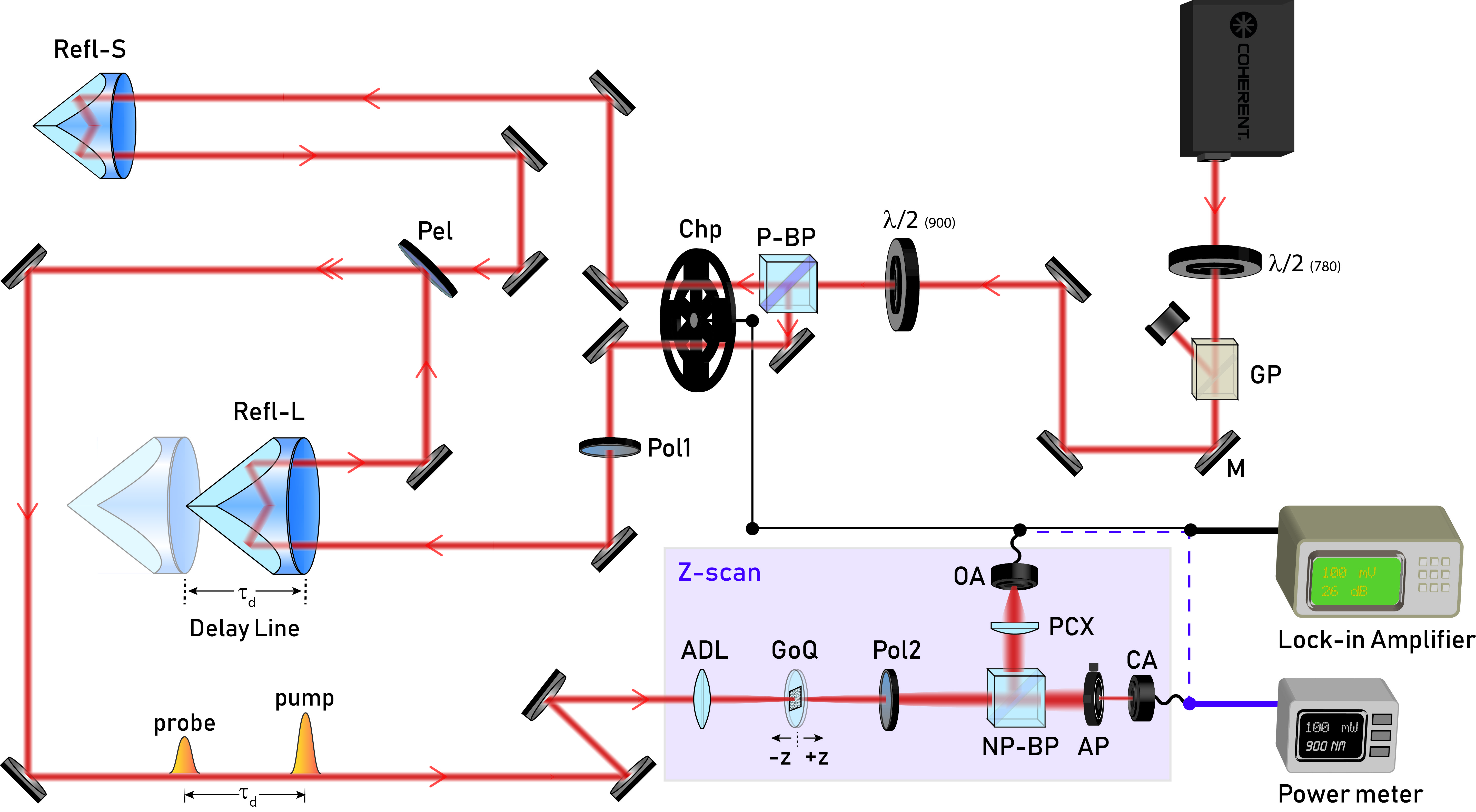}
\caption[Pump-probe integrated Z-scan] {\textbf{Pump-probe integrated Z-scan set-up (PPZS).} The source is Ti:sapphire tunable laser capable of emitting from 690 to 1050 nm at a repetition rate of 80 MHz at $\sim$ 100 fs. The laser has a variable high power attenuator on exit comprising of a half-waveplate ($\lambda/2_{(780)}$) and a Glan-Laser polariser (GP). The beam is directed by silver mirrors (M) into another half-waveplate ($\lambda/2_{(900)}$) and polarising beamsplitter (P-BP) to prepare the orthogonal pump (transmitted) and probe (reflected) beams. Both beams are then directed through the dual-frequency chopper (Chp). The pump is directed into a small static retroreflector (Refl-S), while the probe is directed through a polariser (Pol1) and a large retroreflector (Refl-L) that is mounted on a translation stage that allows for one beam to be delayed relative to another. Upon return, both beams are combined at a pellicle beamsplitter (Pel) and guided to the Z-scan set-up. Both beams are then focused using an achromatic doublet lens (ADL) and impinge upong the graphene on quartz sample (GoQ), followed by another polariser (Pol2) which is oriented parallel to Pol1 for extinction of the pump beam. The beam is then bisected in the far-field by an non-polarising beamsplitter (NP-BP) with the reflected arm directed into the Open Aperture (OA) detector and the transmitted arm directed through an adjustable aperture (AP) into the Closed Aperture (CA) detector. For all Z-scan based measurements, single-beam or temporal, the Power Meter is utilised for data acquisition. In pump-probe mode the chopper and Lock-in Amplifier are used for data acquisition.}
\label{fig:PPZSsetup}
\end{figure}

Originally introduced by Sheik-Bahae \emph{et al.} \cite{ZS1}, the Z-scan technique proved to be an experimentally facile yet sensitive method to extract the phase and magnitude of the Kerr coefficient. The optically induced self-refraction is quantified  by relating the phase modulation of the traversing beam to the transmittance in the far-field in Closed Aperture (CA) configuration, while the Open Aperture (OA) configuration captures the effect of absorption. This technique was modified by Wang \emph{et al.} \cite{PPZS} to extract the temporal evolution through the integration of a secondary time delayed beam to obtain the time-resolved Z-scan measurement. The pump-probe integrated Z-scan set-up (PPZS) is schematically illustrated in Figure \ref{fig:PPZSsetup}. The set-up is elucidated upon in the Methods section at the end of the article. The set-up can be operated in multiple modes of measurement. In the single-beam mode, the set-up is a standard Z-scan measurement. In dual-beam mode, with the addition of a lock-in amplifier, chopper and cross-polarisation filtering, the set-up can be used to perform a standard pump-probe measurement for the extraction of relaxation time constants. When used in conjunction with the Z-scan components, the dual-beam mode is used for the time-resolved Z-scan measurements.

\newpage
\section*{Results and Discussion}

\subsection*{Spectral-dependence} 

\begin{figure}[H]
\centering
\includegraphics[scale=0.2]{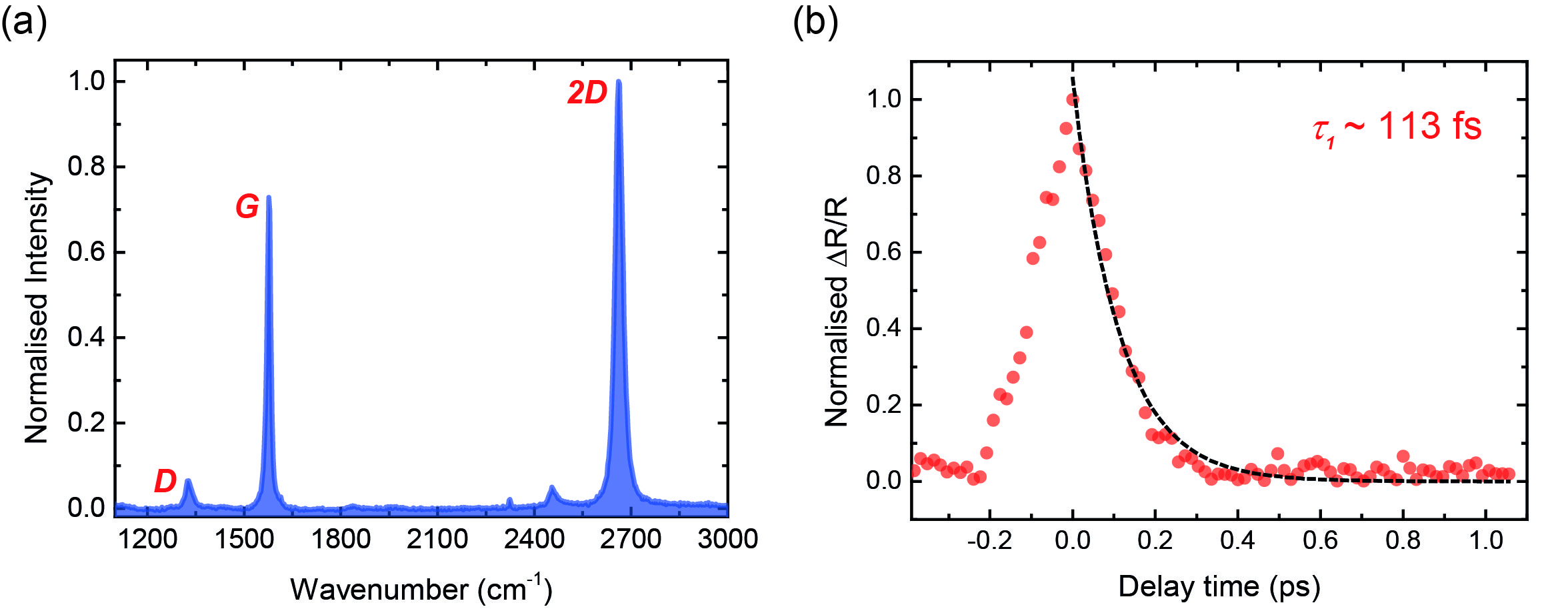}
\caption[Graphene characterisation] {\textbf{Raman and temporal correlation of graphene.} (a) Raman spectra of graphene sample with the G-band, 2D-band and D-band peaks appearing at 1576 cm$^{-1}$, 2661 cm$^{-1}$ and 1328 cm$^{-1}$, confirming monolayer sample. (b) Temporal correlation measurement on graphene with a decay constant of $\tau_1$ $\sim$ 113 fs.}
\label{fig:RPP}
\end{figure}

In order to verify the quality and monolayer nature of the samples, the CVD fabricated graphene on quartz (GoQ) samples (1$\times$1 cm$^2$) were characterised by Raman spectroscopy, shown in Figure \ref{fig:RPP}a. The G-band, 2D-band and D-band peaks appear at 1576 cm$^{-1}$, 2661 cm$^{-1}$ and 1328 cm$^{-1}$, respectively. The relative intensities and widths of the G and 2D peaks confirm that the sample is single layer. We also perform a temporal correlation measurement on the sample, shown in Figure \ref{fig:RPP}b, to obtain a $\tau_1$ relaxation time constant of about $\sim$113 fs.  

\begin{figure}[H]
\centering
\includegraphics[scale=0.2]{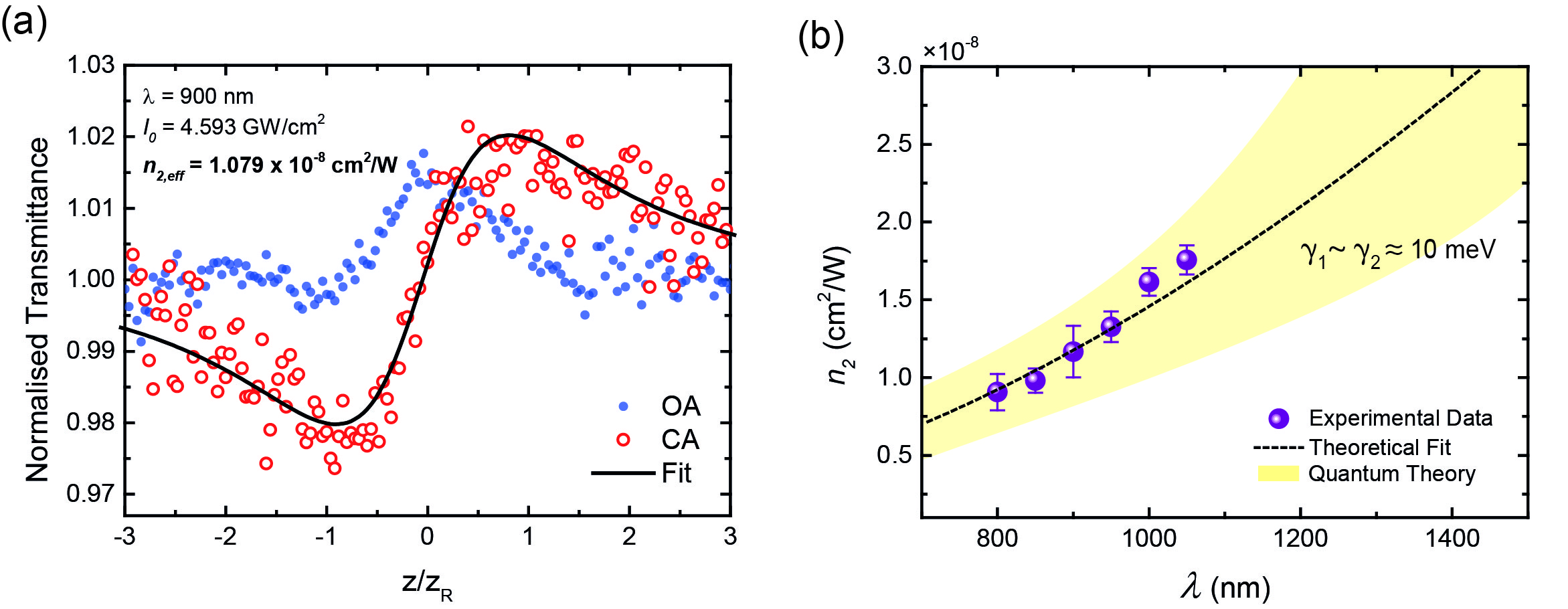}
\caption[Spectral dependence of the $n_{2,eff}$ of graphene] {\textbf{Spectral dependence of the $n_{2,eff}$ of graphene.}  (a) Standard Z-scan trace at 900 nm with the OA, CA and Fit shown. (b) Experimental data obtained using the Z-scan measurement spanning 800-1050 nm with 50 nm increments along with fit obtained from quantum theory. The experimental data matches the $\lambda^2$ dependence predicted in theory. The uncertainties in $\gamma_{1/2}$ are accounted for in our theoretical model by the shaded (yellow) region.}
\label{fig:WlZS2}
\end{figure}

The spectral dependence of $n_{2,eff}$ is obtained in the single-beam Z-scan mode and is measured by scanning the 100 fs excitation source from 800 to 1050 nm with 50 nm increments. The range of the excitation is limited by tunability of the laser (690-1050 nm). Since the measurement is sensitive to the beam quality, the beam is characterised using a knife edge beam profiler for quality and stability over the period of the measurement. To ensure that the Z-scan modulation observed in the far-field is originating from graphene, a bare quartz substrate is measured at 900 nm at high on-axis irradiance to show no effect (see supplementary Figure S5). The on-axis irradiance ranges from 1.5 to 5 GW/cm$^2$, limited by the available power of the laser at each wavelength. A standard acquired Z-scan trace showing the OA, CA and Fit is shown in Figure \ref{fig:WlZS2}a. The effects of saturation resulting in decrease of $n_{2,eff}$ with increasing irradiance is observed and shown at 950 nm in Figure S6 in supporting information. 

\begin{figure}[H]
\centering
\includegraphics[scale=0.20]{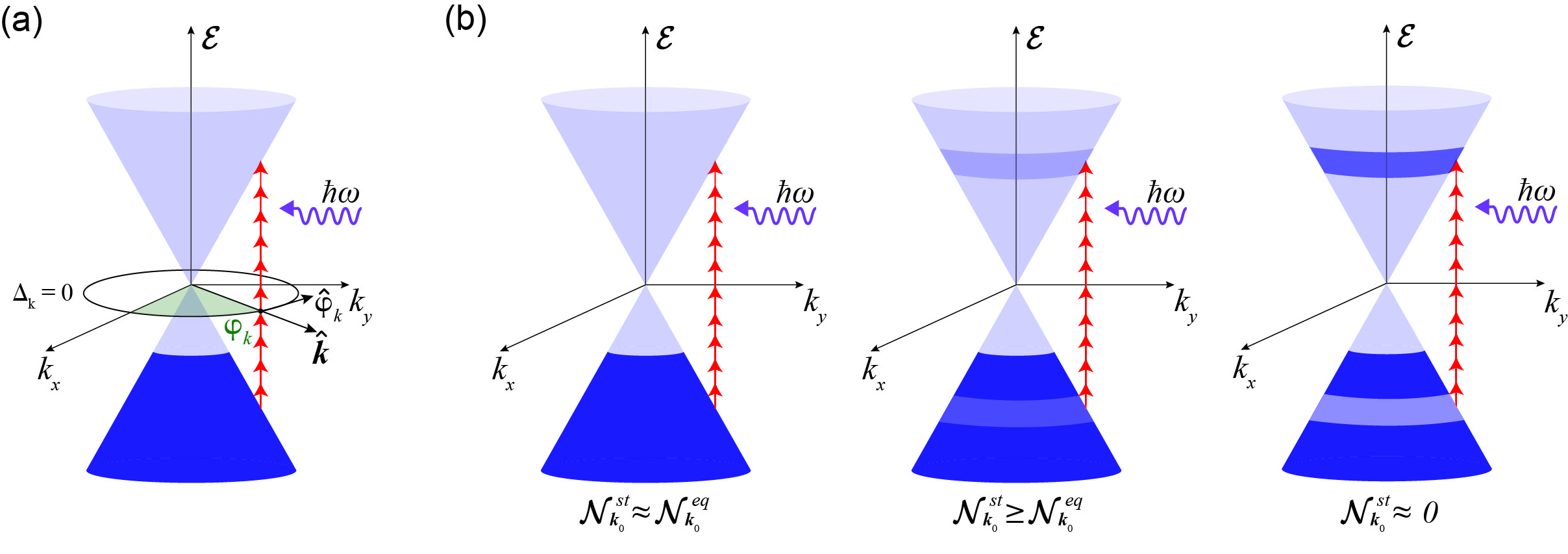}
\caption[Carrier relaxation in graphene leading to Kerr-type nonlinearity] {\textbf{Carrier relaxation in graphene leading to Kerr-type nonlinearity.} (a) Band structure of graphene showing the excitation pulse with energy $\hbar\omega$, the zero detuning circle $\Delta_k = 0$, and $\hat{\varphi}_k$ is a vector in reciprocal space,  (b) the evolution of the steady state population ($\mathcal{N}_{\textbf{k}}^{st}$) and the equilibrium population ($\mathcal{N}_{\textbf{k}}^{eq}$) difference upon intense illumination which disturbs the $\mathcal{N}_{\textbf{k}}^{eq}$ leading to the observation of a Kerr-type nonlinearity in graphene. At very intense illuminations population difference at the zero detuning circle becomes zero and absorption quenching takes place. The colours in the figures represent relative carrier population densities.}
\label{fig:bandstr}
\end{figure}

The cumulative results of the spectral dependence are presented in Figure \ref{fig:WlZS2}b along with the quantum theory based fit. A full set of Z-scan fits from which these results are compiled is also provided for reference (Figure S7 in supporting information). The values for $n_{2,eff}$ in this regime are measured to be positive and range from 9.07$\times$10$^{-9}$ to 1.76$\times$10$^{-8}$ cm$^2$/W in this excitation range which fits well with our quantum model with relaxation coefficients of $\gamma_1 \approx \gamma_2 \approx 10\mathrm{meV}$. The fitting parameters, $\gamma_{1/2}$, are obtained from the time constants derived from the temporal cross-correlation measurement, shown in Figure \ref{fig:RPP}b. However, the pulse width of the pump and probes signals pose some uncertainties on the obtained decay constant of $\tau_1\sim$113 fs as the correlated pulses are of similar pulse durations. This uncertainty is accounted for in our theoretical model and is illustrated by the shaded (yellow) region in Figure \ref{fig:WlZS2}b. Shorter wavelengths are observed to have a $n_{2,eff}$ value that is lower as compared to longer wavelengths, with approximately the same on-axis irradiance. Our calculation suggests a quadratic dependence of $n_{2,eff}$ on $\lambda$ which agrees with the experimental results.  It is noted that $n_{2,eff}$ exhibits negligible dependence on the Fermi level for a low-doped graphene monolayer. In our case $n_{2,eff}$ is dominated by contributions from interband transitions and since the $\lambda^2$ dependence is a direct consequence of the linear band diagram around the Dirac point, the most relevant transition occur at the zero detuning region where $\omega = 2|\textbf{k}|v_F$. Detuning is defined as $\Delta_k = \hbar\omega - \mathcal{E}_{cv}$, where $\omega$ is the frequency of the excitation photon and $\mathcal{E}_{cv}$ is the energy of the transition, see Figure \ref{fig:bandstr}a. The theoretical model employed in our theory uses semiconductor Bloch equations (SBEs) to describe the cooperative intra-interband dynamics of the population difference $\mathcal{N}(\textbf{k},t)$ (between the valence and conduction bands) and the polarisation (coherence) $\mathcal{P}(\textbf{k},t)$ for the Bloch state \textbf{k}. The phenomenological relaxation coefficients, $\gamma_{1/2}$, account for the collective broadening effects for the population and coherence decay, respectively.  Under our theory, the electromagnetic coupling for normal illumination is defined by ${\Phi}_k = \frac{\textbf{E}\cdot\hat{\varphi}_k}{\hbar k}$, where \textbf{E} is the electric field and the unit vector $\hat{\varphi}_k$ is defined as $\hat{\varphi}_k = \hat{z} \times \textbf{k}/k$, shown in Figure \ref{fig:bandstr}a.

The nonlinear optical absorption has multiple origins namely bleaching effects culminating in saturation, and two photon absorption. Since the graphene sample used in our experiment is low-doped (i.e $\mu \ll \hbar\omega$), absorption bleaching due to optically induced Pauli-blocking plays the leading role. This assumption is supported by observation of the absorption drops upon high optical irradiance, refer to the OA trace in Figure \ref{fig:WlZS2}a. Furthermore, since the Fermi energy level is small compared to the energy of the photons, it naturally follows that Pauli blocking around zero detuning region is widely demolished. Under high intensity illumination in a Kerr-type material, the charged carriers undergo ultrafast Rabi oscillations, much faster than their relaxation rate. As a consequence the \textit{steady state population}, $\mathcal{N}_{\textbf{k}}^{st}$, is significantly modified by light, disturbing the distribution at equilibrium, $\mathcal{N}_{\textbf{k}}^{eq}$. The origin of the Kerr-type nonlinearity is the optically induced change to the steady-state population difference. The relaxation dynamics as the population difference evolves is schematically shown in Figure \ref{fig:bandstr}b.  Before saturation takes place, the nonlinear contribution of the field to the population difference around the zero detuning region is a quadratic function of the field magnitude, $\mathcal{N}_{\textbf{k}}^{st} - \mathcal{N}_{\textbf{k}}^{eq} \approx -\frac{1}{\gamma_1 \gamma_2}\mathcal{N}_{\textbf{k}}^{eq}|\Phi_{\textbf{k}}|^2$. The induced nonlinear current oscillating at the frequency $\omega$ is then given by $\mathbf{J}_{NL} = \sum_{\textbf{k}}\hat{\varphi}_k(\mathcal{N}_{\textbf{k}}^{st} - \mathcal{N}_{\textbf{k}}^{eq})\mathcal{L}_{\textbf{k}}(\omega)\Phi_{\textbf{k}}$, where the Lorenztian $\mathcal{L}_{\textbf{k}}(\omega) \triangleq 1/(\gamma_2 + i\Delta_{\textbf{k}})$ accounts for the interband transitions. The nonlinear current $\mathbf{J}_{NL}$ is the microscopic origin of the intensity dependence of the refractive index  so that $n_2  \propto |\mathbf{J}_{NL}|/|\mathbf{E}|^3$. For a small enough $\gamma_2$, the Kerr-type nonlinear induced current is given by
\begin{equation}
\mathbf{J}_{NL} \sim \beta \frac{e^2}{\hbar} g_s g_v D \frac{1}{\gamma_1\gamma_2} |\frac{e}{\hbar k}\textbf{E}|^2 \mathcal{N}^{eq}_{ \mathbf{k}}| _{\Delta_{ \mathbf{k} }=0} \mathbf{E}
\label{eq:nonlincurrent}
\end{equation}
\noindent
where $g_s$ and $g_v$ are the spin and valley degeneracy factors, respectively, $D = 1/4\pi^2$ is the density of states, and $\beta \sim \pi$ is a dimensionless quantity that arises from angular integration around the Dirac cone. Due to low doping, $\mathcal{N}^{eq}_{\textbf{k}} \approx 1$ over the zero detuning circle. Therefore, the frequency dependence of nonlinear current is dominated by the $1/k^2$ term appearing in Eq. \ref{eq:nonlincurrent}. Over the $\Delta_k=0$ circle the Bloch wave-number $k$ is linearly proportional to the frequency which in turn yields $\lambda^2$ dependence of the Kerr nonlinear coefficient.  

It is worth pointing out that the quadratic  wavelength dependence of the Kerr coefficient is a direct consequence of the linear energy-momentum dispersion of the Dirac quasi-particles. The calculations outlined above can be effectively applied to all two-level condensed matter systems by appropriately replacing interband coupling into Eq. \ref{eq:nonlincurrent}. However, exclusive to graphene is the $1/k$ dependence of the interband coupling $\Phi_{\mathbf{k}}$ which results in the square wavelength dependence of the Kerr coefficient .    

\newpage
\subsection*{Temporal evolution}

\begin{figure}[H]
\centering
\includegraphics[scale=0.18]{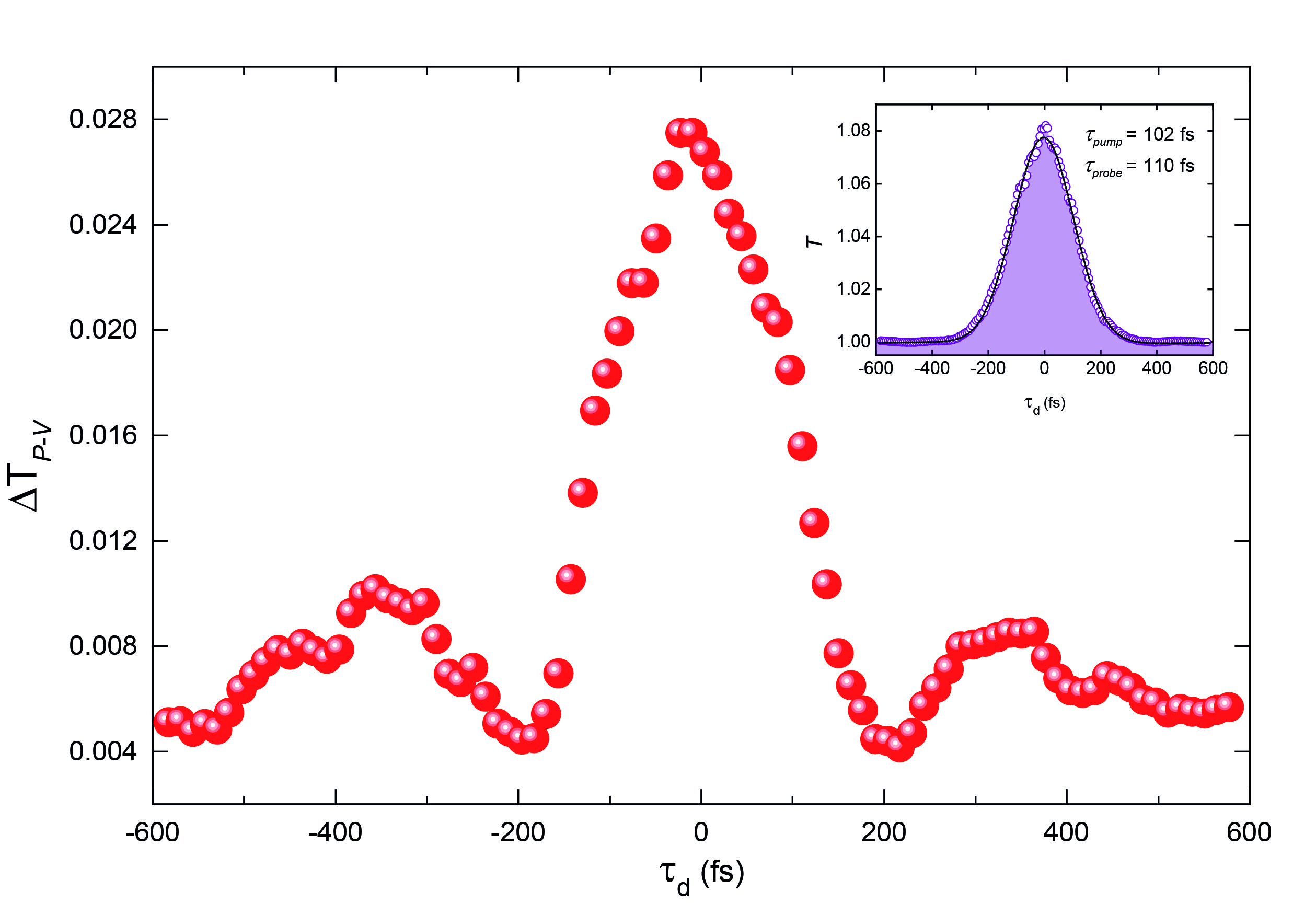}
\caption[Pump-probe integrated Z-scan measurement] {\textbf{Pump-probe integrated Z-scan measurement.} Time-resolved degenerate Z-scan measurement of graphene. Since $\Delta T_{pv}$ is proportional to $n_{2,eff}$, the plot follows the evolution of the induced phase modulation in graphene. The effect peaks at zero time delay and relaxes at longer time scales, showing some oscillatory behaviour as it relaxes. \textbf{(Inset)} Temporal cross correlation of pump and probe pulses at peak position.}
\label{fig:PPZS}
\end{figure}

The temporal evolution of the nonlinearity is obtained using the dual-mode PPZS set-up. A single-beam (pump) Z-scan measurement is performed to locate the peak and valley positions of the sample. In order to locate the zero delay position of the probe, the sample (GoQ) is placed at the peak position and the probe is scanned until a cross-correlation signal of the pump and probe pulses is obtained, shown in the inset of Figure \ref{fig:PPZS}. The FWHM of the signal is 150 fs, which gives a pulse duration of 110 fs for the probe, considering a pulse duration of 102 fs for the pump. A similar scan is run at the valley position and both time-resolved data sets are used to extract the differential peak-valley transmittance, $\Delta T_{pv}(t_d) = \pm \Big[\frac{T(t_d, Z_p)}{T_{OA}(t_d,Z_p)} - \frac{T(t_d,Z_v)}{T_{OA}(t_d,Z_p)}\Big]$, where $Z_{p/v}$ are the positions of the peak and valley, respectively, $t_d$ is the probe delay, and $T_{OA}$ is the OA transmittance for normalisation. The sign of $\Delta T_{pv}$ is given by the sign of $Z_p-Z_v$. The pump power used for this measurement was 200 mW, equating to an intensity of $\sim$ 3.5 GW/cm$^2$. The pump to probe power ratio was kept at 20:1. In general, the far-field aperture should only allow about 1$\%$ of light transmittance to isolate the effect of wave front distortion due to phase modulation, however, due to the weak signal of the probe the aperture is opened to allow 10$\%$ of transmittance with additional averaging at each acquisition point. 

The temporal evolution of the nonlinearity is shown in Figure \ref{fig:PPZS}b. The effect peaks at zero probe delay and relaxes on the time-scale of $\tau_1$, while at longer time scales, $\tau_2$, $\Delta T_{pv}$ shows no discernible variation. The symmetric shape of the figure leads us to believe that in this measurement we are simply observing the probe following the pump signals rather than unravelling the nonlinear phenomena hidden at shorter time scales. This is also supported by the fact that the pulse duration of the pump and probe is similar to the measured temporal cross-correlation extracted in Figure \ref{fig:RPP}b. Therefore, we can conclude that the relaxation dynamics contributing to the observation of the nonlinear refraction are simply too fast to be measured by our laser pulses in this manner. The variation of $\Delta T_{P-V}$ or the induced phase shift $\Delta \Phi$ seen in Figure \ref{fig:PPZS} relates to the Gaussian power distribution within the pump pulse which reaches its maximum value (i.e. $\sim$ 3.5 GW/cm$^2$) at the center peak of the pulse. Therefore, according to $n_{2,eff} = \frac{\sqrt{2}\Delta\Phi_0}{k_0 I_0 L_{eff}}$ (See Methods), the $n_{2,eff}$ for the peak power is extracted to be 1.12$\times$10$^{-8}$ cm$^2$/W. As the intensity decreases moving away from the peak, the $n_{2,eff}$ increasing in accordance with saturation. This measurement clearly shows that on the time scale greater than the heating and cooling times, the probe follows the pump and Kerr nonlinearity is tunable and controllable over this time scale. The modulation is understood through the relation $n = n_0 + \Delta n$, where $\Delta n = n_{2,eff}I$, and the $\Delta n$ parameter is modified as the $n_{2,eff}$ varies nonlinearly on the intensity prior to complete saturation \cite{zhang_2012}. 

For the purposes of all-optical switching the on/off time of the nonlinearity is controlled by the pulse-duration and power of pulse. If the pulse-duration is longer than the effective relaxation times of the nonlinearity, then the evolution of the observed nonlinearity simply follows the Gaussian power distribution of the pulse. In this way through modulation of the $\Delta n$ term, by changing the pulse properties, the switch can be controlled. However, there is saturation that takes place at relatively high intensities, leading to a deviation between the pump power within the pulse and the observed $\Delta n$.

\newpage
\subsection*{Pulse-width dependence}

\begin{figure}[H]
\centering
\includegraphics[width=\linewidth]{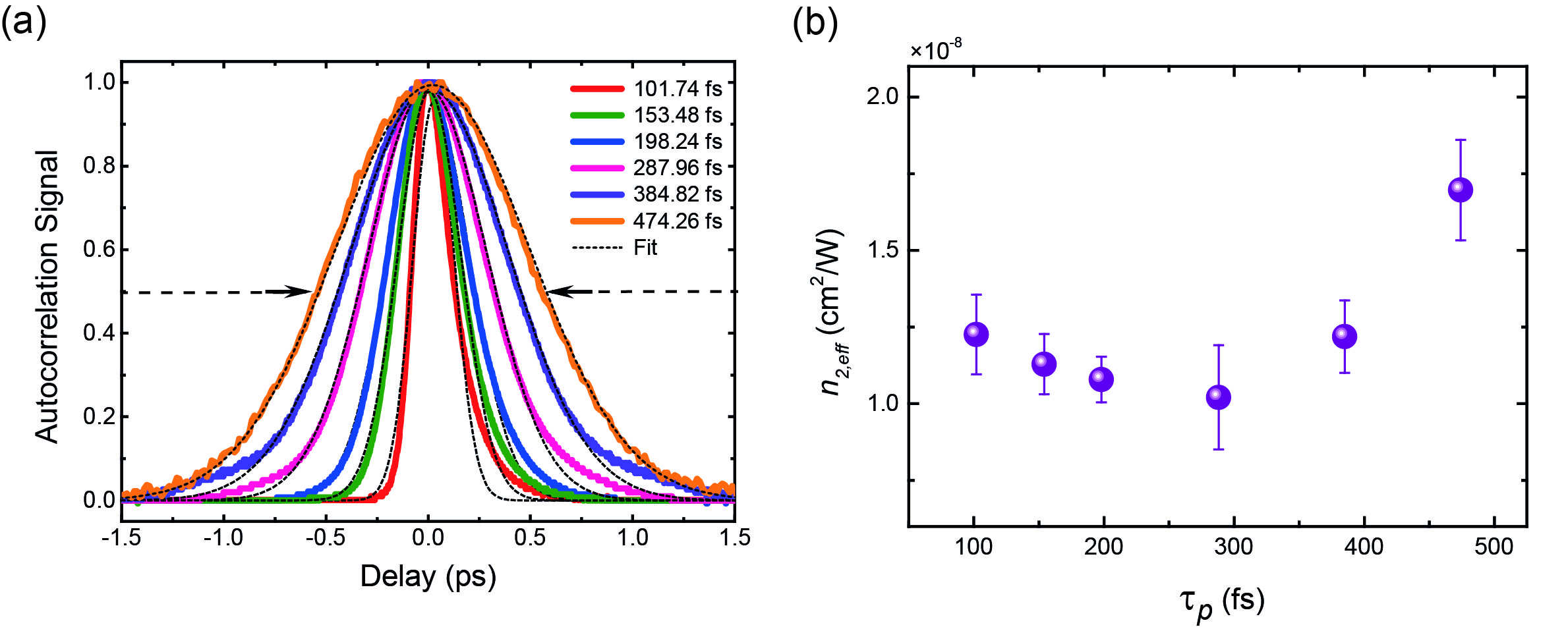}
\caption[Pulse-width dependence of graphene] {\textbf{Pulse-width dependence of $n_{2,eff}$.} (a) Autocorrelation traces and fits for the temporally stretched beam at 900 nm. (b) The dependence of $n_{2,eff}$ on the excitation pulse-width. The $n_{2,eff}$ is observed to increase as the pulse-width becomes longer.}
\label{fig:ps}
\end{figure}

It is well understood that the relative timescales of the excitation pulse and the system response times determine the induced dynamics in the system. To investigate the effects of pulse duration on nonlinear refraction, we stretch the $\sim$ 100 fs pulse upto $\sim$475 fs at 900 nm to observe the effects in the long pulse regime; longer than the duration of the nonlinearity or relaxation time, $\tau_1$.  Pulse stretching is achieved using a dispersion based prism-pair apparatus, where the pulse can be temporally expanded by varying the separation distance of prisms\cite{sherriff}. A schematic of the set-up is shown in Figure S8 in supplementary information. In our set-up we use N-SF11 flint glass prisms. The maximum allowed pulse expansion is limited by the space on the optical table. The autocorrelation curves for the expanded pulses at which the Z-scan is performed are given in Figure \ref{fig:ps}a and the results of the pulse-width dependence are provided in Figure \ref{fig:ps}b. A set of Z-scan fits for the presented data is also provided in Figure S9 in supporting information.

The results show a clear dependence of $n_{2,eff}$ on the excitation pulse duration, with $n_{2,eff}$ becoming larger with increasing pulse-width. This trend was theoretically speculated by Vermeulen \emph{et al.}\cite{vermeulen_spm} and shown here experimentally. The $n_{2,eff}$ ranges from 1.02$\times$10$^{-8}$ to 1.7$\times$10$^{-8}$ cm$^2$/W in this pulse duration regime with a minimum between 200-300 fs. The decay constant $\tau_1 \sim$ 113 fs measured in Figure \ref{fig:RPP}b falls near the beginning of the data set. However, doping can modify the decay constant values with \textit{p}- and \textit{n}-doping making the time constants long and shorter, respectively\cite{johan}. This data trend coincides with previously reported Z-scan studies on graphene performed with picosecond excitation where the $n_{2,eff}$ is larger than what is reported in the femtosecond regime. In addition to this, a similar comparative analysis performed on carbon disulphide (CS$_2$), reference material used for calibration of Z-scan measurements, revealed a similar dependence of the $n_{2,eff}$ on pulse duration\cite{niko}. In a recent publication\cite{vermeulen_spm}, this pulse-width dependence of $n_{2,eff}$ is theoretically derived for a regime when the effective decay constant is larger than the pulse duration in Z-scan measurements. The interplay of relative carrier heating and cooling times is said to induce a nonlinear response that may not originate only from the conventional electronic Kerr-type nonlinearity but also from what they refer to as saturable photoexcited-carrier refraction (SPCR). The saturability in graphene deviates from saturability in other 2D materials due to the presence of its unique gapless band structure which facilitates spontaneous saturation near the Dirac point even when there is no field \cite{semnani18}. Therefore, it is apt that we refer to the Kerr-type nonlinearity characterised using the Z-scan method as $n_{2,eff}$.

\section*{Conclusions}

Through the systematic measurement and analysis of the effect of the spectral and temporal properties of the pulse, we were able to gain a more fundamental understanding of the parameters governing the observed nonlinear optical effect in graphene. This sheds significant light on the widespread debate in the field regarding the large variation in the $n_{2,eff}$ value for graphene due to varying experimental conditions and sample preparation techniques. The dependence of $n_{2,eff}$ on the exciting wavelength revealed a quadratic ($n_{2,eff} \propto \lambda^2$) relationship both experimentally and theoretically. The time-resolved Z-scan measurement revealed that the heating and cooling dynamics within our graphene sample are too fast to be probed using $\sim$ 100 fs pulses using this method. However, this method reveals a practical application in all-optical ultrafast switching of the nonlinearity, completely controlled by the pulse-duration and power of the impinging laser pulse. Through the pulse-duration dependent measurement, we were able to confirm the theoretically predicted relationship between $n_{2,eff}$ and the laser pulse-duration in the hundreds of femtosecond regime, with $n_{2,eff}$ growing larger with longer pulse-duration. Throughout all our experiments, the value for $n_{2,eff}$ remains positive. With this study we have gained a fundamental understanding of the underlying processes governing the nonlinear optical phenomena and parameters to modulate the effect. In doing so we can accurately form predictable models for the purposes of nonlinear photonic device design.

\section*{Methods}

\paragraph{Pump-probe integrated Z-scan measurement.} The experimental set-up for this measurement is illustrated in Figure \ref{fig:PPZSsetup}. The excitation source is a high-power Ti:sapphire laser (Coherent Chameleon Vision S) delivering $\sim$100 fs pulses at wavelengths tunable from 690 to 1050 nm, at a repetition rate of 80 MHz. When operating in single beam mode, the set-up is a classic Z-scan measurement. The beam is focused onto the sample using an anti-reflection (AR) coated achromatic doublet lens (ADL) with a focal length of 75 mm. The beam is bisected in the far-field by an AR coated non-polarising beam splitter (NP-BP) to obtain the OA and CA profiles. The OA is used to normalise for absorption and laser fluctuations in the CA trace. The normalised CA transmittance is fitted to the equation: $T(x,\Delta\Phi_0) \simeq 1 - \frac{4\Delta\Phi_0x}{(x^2 + 9)(x^2 + 1)}$, where $x$ is the Rayleigh length ($z_R$) normalised position ($z/z_R$) and $\Delta\Phi_0$ is the nonlinearity induced phase shift. When a pulsed source is used with pulse-width comparable to the duration of the nonlineairty, the nonlinear refractive index is extracted via  $n_2 = \frac{\sqrt{2}\Delta\Phi_0}{k_0 I_0 L_{eff}}$, where $k_0$ is the wave vector, $I_0$ is the on-axis irradiance and $L_{eff}$ is the effective length of the sample \cite{ZS2}. In dual-beam mode, we use degenerate pump and probe at 900 nm, in a collinear configuration with a 20:1 pump/probe ratio. The pump and probe are orthogonally polarised using a $\lambda/2$ waveplate ($\lambda$/2), polarising beam-splitter (P-BS) and polariser (Pol1) before the sample. An analyser polariser (Pol2) placed after the sample is rotated to achieve extinction of the pump beam.

\bibliography{main}

\section*{Acknowledgements}

S.T., B.S. and A.H.M. acknowledge the financial support received from the Natural Science and Engineering Research Council of Canada (NSERC) and internal funding from U.W. B.S. and S.S.N. also acknowledge the financial support provided by the Quantum Quest Seed Fund (QQSF).

\section*{Author contributions statement}
S.T., B.S., and A.H.M. defined the project and conceived the experiment(s). S.T. and B.S. conducted the experiment(s), B.S. developed the theoretical model and S.T. analyzed the results, under the guidance of A.H.M. and S.S.N. S.T. and B.S. wrote the manuscript and all authors reviewed the manuscript.
\section*{Additional information}

\textbf{Competing interests}:
The author(s) declare no competing interests
\newline
\textbf{Supplementary information} is available from the corresponding author upon request.
\newline
\textbf{Data availability:} Experimental data presented in this article is available from the corresponding author upon request.

\end{document}